\documentclass[12pt]{article}
\usepackage{amsmath}
\usepackage{graphicx,psfrag,epsf}
\usepackage{enumerate}
\usepackage{enumitem}   
\usepackage{caption}
\usepackage{placeins}
\usepackage[hidelinks,colorlinks=true,allcolors=blue]{hyperref}
\usepackage{subcaption}
\usepackage{amsfonts}
\usepackage{url} 
\usepackage[natbibapa]{apacite}
\usepackage{setspace}
\newcommand{\blind}{0}
\usepackage[margin=1in]{geometry}

\begin{document}

\def\spacingset#1{\renewcommand{\baselinestretch}%
{#1}\small\normalsize} \spacingset{1}


\if1\blind
{
  \title{\bf A non-parametric approach to detect patterns in binary sequences}
  \author{Author 1\thanks{
    The authors gratefully acknowledge \textit{please remember to list all relevant funding sources in the unblinded version}}\hspace{.2cm}\\
    Department of YYY, University of XXX\\
    and \\
    Author 2 \\
    Department of ZZZ, University of WWW}
  \maketitle
} \fi

\if0\blind
{
  \bigskip
  \bigskip
  \bigskip
  \begin{center}
    {\LARGE\bf A Non-Parametric Approach to Detect Patterns in Binary Sequences}\\

   {\large Anushka De \\ Indian Statistical Institute, Kolkata\\anushka.isical@gmail.com}
\end{center}
  \medskip
} \fi

\bigskip
\doublespacing
\begin{abstract}
In many circumstances, given an ordered sequence of one or more types of elements or symbols, the objective is to determine the existence of any randomness in the occurrence of one specific element, say \textit{type 1}. This method can help detect non-random patterns, such as wins or losses in a series of games.  
Existing methods of tests based on total number of runs or tests based on length of longest run (Mosteller (1941)) can be used for testing the null hypothesis of randomness in the entire sequence, and not a specific type of element. Moreover, the Runs Test often yields results that contradict the patterns visualized in graphs showing, for instance, win proportions over time. This paper develops a test approach to address this problem by computing the gaps between two consecutive type 1 elements, by identifying patterns in occurrence and directional trends (increasing, decreasing, or constant), applies the exact Binomial test, Kendall's Tau, and the Siegel-Tukey test for scale problems. Further modifications suggested by Jan Vegelius(1982) have been applied in the Siegel Tukey test to adjust for tied ranks and achieve more accurate results. This approach is distribution-free and suitable for small sample sizes. Also comparisons with the conventional runs test demonstrates the superiority of the proposed method under the null hypothesis of randomness in the occurrence of \textit{type 1} elements. 
\end{abstract}

\noindent%

\newpage
\section{Introduction}
\label{sec:intro}
In many cases, when given an ordered sequence of one or more types of elements or symbols, the objective is to determine any existence of non-randomness in the occurrence of one of the elements. Consider a series of games played by a single player; our task is to assess the existence of any non-random pattern specifically in their wins or losses. This problem essentially reduces to determining the randomness of a single element type in an ordered dichotomous sequence\\ 
 The conventional Runs test  (\cite{gibbons}) is one of the most well-known and easiest-to-apply tests for randomness in a sequence of observations. A \textit{run} is defined to be a succession of one or more types of symbols, followed and preceded by a different symbol or no symbol at all. Clues indicating a lack of randomness are provided by any tendency of the symbols to exhibit a definite pattern in the sequence.
 Both the number and length of the runs should reflect the existence of some sort of pattern. When the alternative is non-randomness, a test based on the total number of runs—whether too few or too many—suggests a lack of randomness. \\
 Assume an ordered sequence of $n$ elements of two types, $n_1$ of the first type and $n_2$ of the second type, where $n_1+n_2=n$. If $R_1$ is the number of runs of type 1 elements and $R_2$ is the number of runs of type 2 elements, the total number of runs in the sequence is the random variable $R=R_1+R_2$. In order to derive a test for randomness based on $R$, the Runs test computes the probability distribution of $R$ when the null hypothesis of randomness is true. This test can be both-sided or one-sided. The alternative can be simply non-randomness or trend. Since the presence of a trend would usually be indicated by a clustering of like objects, which is reflected  by an unusually small number of runs, a one-sided test is more appropriate for trend alternatives. The Runs Test is only one way of using information about runs to detect patterns in an arrangement. Other statistics of interest are provided by considering the lengths of these runs. Since a run that is unusually long reflects a tendency for like objects to cluster and therefore possibly a trend, \cite{mosteller} suggested a test for randomness based on the length of the longest run. Both tests use only a portion of information available, since the total number of runs, although affected by the lengths of the runs, does not directly makes use of information regarding these lengths and the length of the longest run only partially reflects both the lengths of other runs and the total number of runs. Furthermore, considering the objective of this paper's test, where the null hypothesis is randomness in the occurrence of, say, type 1 elements and the alternative is a definite pattern in these elements, the aforementioned tests fail in purpose. These test based on total number of runs or length of longest run consider all the type of elements in determining the test statistic. Intuitively, if the longest run in a binary sequence of games is of losses  and we need to determine any pattern in win occurrence the test based on length of longest run is not suitable. \\
This paper differs from the aforementioned studies in many respects. A distribution-free approach is proposed, focusing on testing the null hypothesis of randomness in the occurrence of a single type of element in a binary or dichotomous sequence. This technique is based on the idea that a non-random pattern will exhibit either a defined pattern in the occurrence of type 1 element \textit{(say)}, or a directional trend, like increasing, decreasing or constancy in the occurrence of the type 1 element. 
 This study makes the use of positions of occurrence of a type of element in the sequence and illustrates further on the idea of lag in the consecutive occurrences. This approach uses the exact Binomial test, Kendall's $\tau$ and the Siegel-Tukey test (\cite{kendall-tau}, \cite{siegel-tukey}) to reach a conclusion. 
 Our approach does not impose any distributional constraints and so is non-parametric in nature. \\
 The remainder of this paper is organized as follows. Section 2 illustrates on generation of a binary sequence \textit{(consisting of 0 and 1)} which apparently will be non-random, plotting a graph of the consequent proportion of \textit{ones} and comparing the visualized patterns from the graphs with the results of the conventional Runs test on such sequences. The objective of this simulation is to motivate the reader into looking for newer methods for addressing the problem.  Section 3 highlights the rationale behind the proposed approach, the outline of the approach and simulation results of comparison of the proposed test with the ordinary test is presented. Further the adjustment for ties in the Siegel-Tukey test as proposed by \cite{vegelius} has been incorporated in our approach to achieve a more accurate result. Section 4 concludes the paper. 
\section{The Conventional Runs Test}
\label{sec:runstest}
\subsection{Generating a binary sequence with differential rate}
\label{Generate: binary_seq}
To test the null hypothesis of randomness in the occurrence of \textit{ones} in a binary sequence of \textit{zeros} and \textit{ones}, we first simulate a binary sequence where the probability of \textit{ones} increases with each subsequent position.  
For ease of interpretation, the binary sequence symbolises the outcomes of various games played by a player, with \textit{wins} represented by $1$ and \textit{losses} represented by $0$.  
This simulation process is modelled as $$P[X=1|i^{th} game]=a+b(i-c),$$ where $1\leq i\leq n$, $n$ denotes the length of the dichotomous sequence (total number of games), $b$ $(0<b<1)$ is the scale factor that generates an increasing sequence of wins (ones) over games played in the simulations, the intercept term $a$ $(0<a<1)$ is kept at $a=0.5$. An additional term $c$ to balance fluctuations ($0<c<\frac{a}{b}+n$).

\begin{figure}[h!]
    \centering
    \includegraphics[width=1\textwidth]{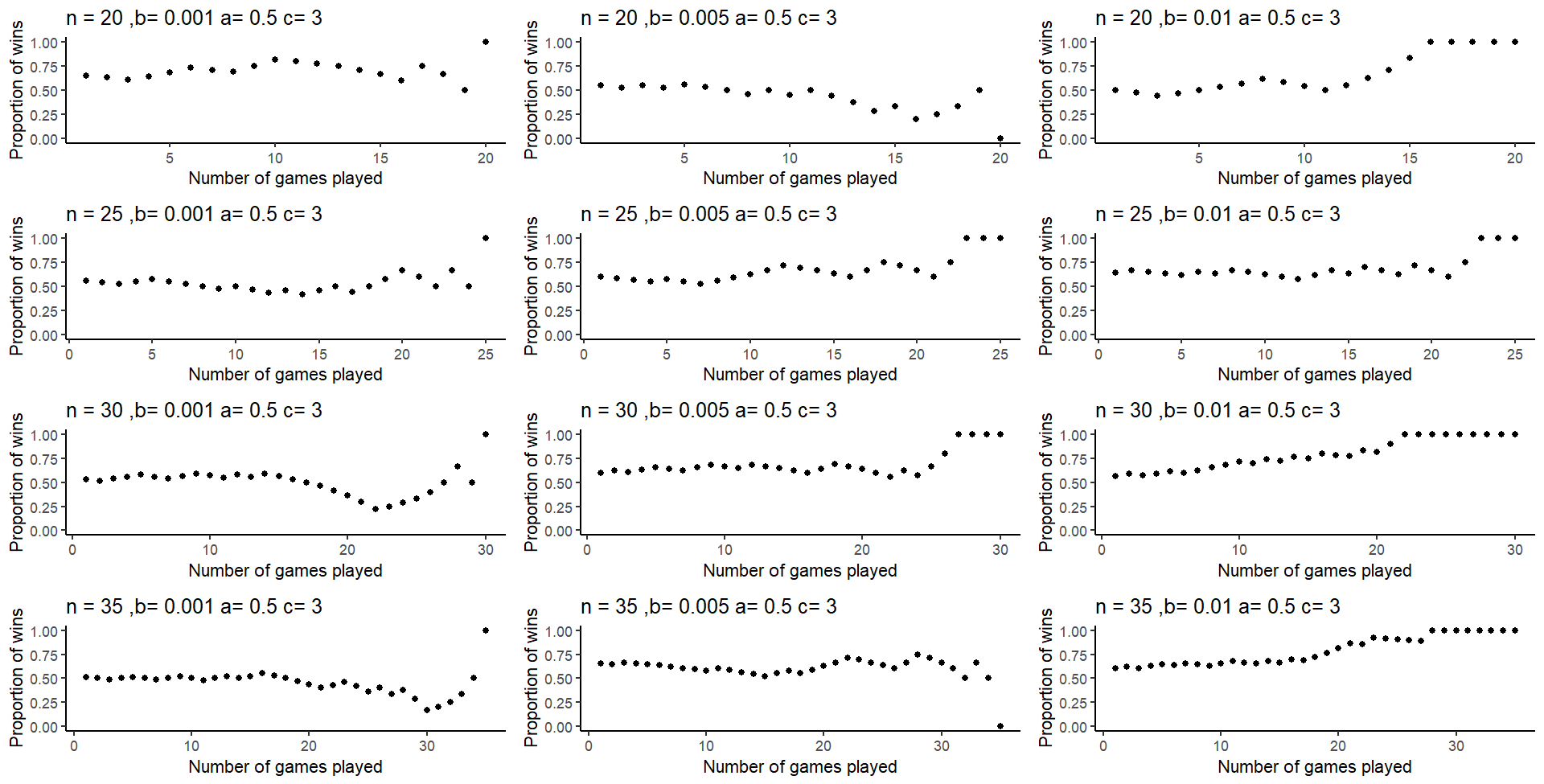}
    \caption{Win proportion vs. number of games played for the 12 combinations of {\it (n,b)}}
    \label{fig:runtest-results}
\end{figure}

We simulated a binary sequence with possible ``increasing trend" with different values of n (20, 25, 30, 35) and different values of $b$ (0.001, 0.005, 0.01) and ran the Runs test using the existing libraries
in R on each of the 12 combinations. 
The win proportion is plotted with the y-axis showing the ratio of games won after playing $(i-1)$ games and the x-axis indicating the game number. The following figures illustrate the win proportion in the 12 binary patterns (Figure \ref{fig:runtest-results}) for visual insight.  

\FloatBarrier
\subsection{Results from the Runs Test}
We observed that for all 12 combinations the Runs Test results were insignificant when the alternative hypothesis was non-randomness or trend, meaning the null hypotheses of randomness holds true. However, close observation of the graphs reveals non random patterns (increasing or constant direction). \\
\\ 
\textbf{An Illustrative Example:}\\
\label{sec:run_test_illustrative_example}
Consider the binary sequence of length 25, representing the outcomes of 25 rounds of a game (with \textit{wins} represented by 1 and \textit{losses} represented by 0): 
$$S_n = \{ 1, 0, 1, 1, 0, 0, 0, 0, 1, 1, 1, 0, 1, 0, 0, 1, 1, 1, 0, 1, 1, 1, 1, 1, 1 \},$$ which is ordered, i.e., every outcome occurs at consecutive time steps. \\



\noindent \textbf{Result of the Runs Test executed on the above sequence:} \\
statistic = -0.67642, runs = 11, $n_1$ = 16, $n_2$ = 9, $n$ = 25, p-value = 0.4988 \\

\begin{figure}[h!]
    \centering
    \includegraphics[width=0.5\textwidth]{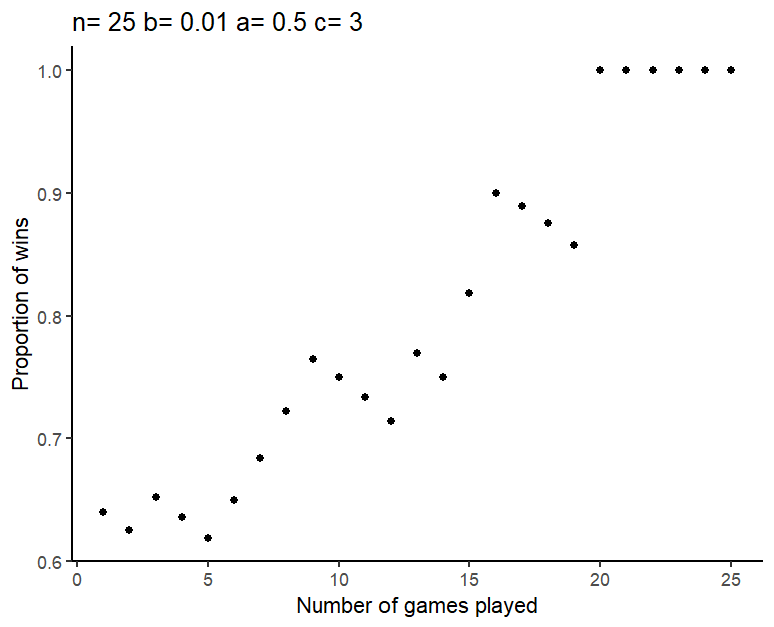}
    \caption{}
    \label{fig:runtest-example1}
\end{figure}

The win proportion for this sequence is plotted in Figure \ref{fig:runtest-example1}. 
\begin{table}[hbt!]
\centering
\scalebox{0.9}{
\begin{tabular}{ c|c c ccccccccc } 
 \hline
Game No. & 1& 2& 3& 4& 5& $\hdots$& 21& 22& 23& 24& 25 \\
Outcome &1&0&1&1&0& $\hdots$& 1& 1& 1& 1&1\\
Win proportion & $\frac{16}{25}$&$\frac{15}{24}$&$\frac{15}{23}$&$\frac{14}{22}$&$\frac{13}{21}$& $\hdots$& $\frac{5}{5}$& $\frac{4}{4}$& $\frac{3}{3}$& $\frac{2}{2}$&$\frac{1}{1}$\\
 \hline
\end{tabular}}
\caption*{Illustration on the calculation of win proportion used in formulation of the graphs}
\label{tab:graph_formation}
\end{table}

\FloatBarrier
\FloatBarrier
We observe a somewhat increasing pattern of win proportion. Also note that after a certain stage (game number 19 onward) the proportion becomes constant and equal to 1. Moreover the sequence does not display a random pattern. Note that the Runs Test considers the number of runs of objects of a particular type.  So in the example, from game number 9 onward, the frequency of ones is more (13/17). However, when computing the number of runs, having fewer zeros affects the entire Runs Test statistic, thereby increasing the total number of runs. 

In the following sections we describe our proposed approach and show results obtained from large number of simulations and its comparison with the Runs Test. 
\section{Proposed Test}
\subsection{ Rationale behind the Proposed Approach:}
\begin{enumerate}
    \item 
   As a first step, this approach considers the elements with the higher frequency in the sample.
Consider the binary sequence: $1,0,1,1,1,0,0,0,1,1,0,1,1$, where the \textit{total number of elements} is $n =13$, and the element $1$ has higher frequency $n_1 =8$. 
Then the positions of these elements will be considered. So $y = (1,3,4,5,9,10,12,13)$ and the corresponding lag vector will be $y' = (3-1,4-3,5-4,9-5,10-9,12-10,13-12)=(2,1,1,4,1,2,1)$. The method is so developed so that it can used for small samples. Thus we utilise the maximum possible information by considering the element with larger frequency. So if in series of games played, a player lost more games than he won, and the nature of his wins need to be assessed, then a fair analysis on the occurrence of the games lost will be quite indicative. For example, despite losing more games than winning, if we could ascertain that the losing direction is decreasing then it will indicate towards a possible increasing number of wins in future.   
\item  Determining the pattern:
These steps focuses on the \textit{Binomial test component} of the proposed approach. 
If there exists a pattern in the occurrence of \textit{wins} then the \textbf{distances} between any two consecutive \textit{ones}, that is, elements of $y'$ will be relatively ``similar" in general. Thus the elements of lag vector will differ by a somewhat common unit.
Thus the total number of lags, which are less than or equal to the ``mean" will be large.  \textit{A proof of the claim is provided in Appendix A}.  A Binomial Test with null hypothesis of equal proportion will be able to determine this, that is, $H_0: p=0.5$, since under the null the distribution is symmetric about the mean. A significant p-value will indicate existence of some pattern in the occurrence of ones (\textit{wins}). 
\item Determining the direction: 
There can be a possibility that no definite pattern exists in the occurrence of the mostly occurring element but the graphs (like the win-proportion vs number of games played) reflect some information as to the direction of say, \textit{wins}.  
The next component of the test focuses on determining the direction of the most frequent element, if any. Assume that the element 1 (symbolic for \textit{wins}) has higher frequency. To compare the occurrence of these elements over the period of time, and measure the degree of association between the time lag vector ($y'$) and a vector symbolising the increasing time interval say $(1,2,\hdots,(k-1))$ a suitable correlation coefficient must be used. Next note that the vector for time period can consist of any set of elements which are in increasing order. Thus the correlation coefficient should be irrespective of the magnitude of the elements and only consider their order. This paper makes use of the Kendall correlation coefficient \cite{kendall-tau} owing to its interpretation of concordant and discordant pairs, which stays relevant in our problem. Since our occurrence of ties will be common in such applications, we used the Kendall Tau adjusted for ties. 
\par Let $(x_1,y_1),(x_2,y_2),\hdots,(x_n,y_n)$ be a set of observations from random variables $(X,Y)$. Any pair of observations $(x_i,y_i)$ and $(x_j,y_j)$ where $i<j$ are said to be concordant if either $x_i>x_j$ and $y_i>y_j$ or $x_i<x_j$ and $y_i<y_j$, and to be discordant if either $x_i<x_j$ and $y_i>y_j$ or $x_i>x_j$ and $y_i<y_j$ , otherwise the pair is said to be tied. The Kendall-correlation coefficient adjusted for the ties is:  $\frac{n_C-n_D}{\sqrt{n(n-1)-\sum_i^k r_i(r_i-1)}\sqrt{n(n-1)-\sum_i^l s_i(s_i-1)}}$, where $n_C:$ number of concordant pairs, $n_D:$ number of discordant pairs,  $r_1,r_2,\hdots,r_k$ are the lengths of the k ties of X and $s_1,s_2,\hdots,s_l$ are the lengths of the l ties of Y.
Thereafter using hypothesis testing, we can use the Kendall rank coefficient as a test statistic to establish whether two variables (time period and lag) may be regarded as statistically dependent or not. This is a non-parametric test as it does not make any assumptions on the distribution of the two variables. Under the null hypothesis of independence of X and Y, the sampling distribution of $\tau$ has an expected value of zero. By testing the alternative hypotheses $H_{1a}: \tau >0$ and $H_{1b}: \tau <0$, we can draw conclusions about the direction of occurrence of\textit{ ones}. For instance, if occurrence of wins increases over time, then the correlation between distance of the \textit{ones} and the increasing time period, say $(1,2,3,\hdots , (k-1))$ will be negative, and vice versa. 
\par Apart from increasing or decreasing direction in a binary pattern, there can be third plausibility of a constant direction in the occurrence of the element, that is the win proportion remains constant over the period of time. To test for this, the distribution-free measures to test for difference is scale can be used. The non-parametric test proposed by \cite{siegel-tukey} tests for differences in scale between two groups. The test is used to determine if one of two groups of data tends to have more widely dispersed values than the other. In other words, the test determines whether one of the two groups tends to move, sometimes to the right, sometimes to the left, but away from the center (of the ordinal scale).  
The hypothesis for testing is stated as: $H_0 : \sigma^2_A=\sigma^2_B$ and $H_1 : \sigma^2_A >\sigma^2_B$. \\
The test is based on the principle that given two groups A and B with $k_1$ observations for the first group and $k_2$ observations for the second (so there are $n = k_1 + k_2$ total observations). If all $n$ observations are arranged in ascending order, it can be expected that the values of the two groups will be mixed or sorted randomly, if there are no differences between the two groups (following the null hypothesis $H_0$). This would mean that among the ranks of extreme (high and low) scores, there would be similar values from Group A and Group B. 
\par The Siegel-Tukey test statistic is defined such that it is sensitive to differences in scale using a rearrangement of the first $n$ positive integers as weights. The weights for $n$ even are: 
\begin{table}[hbt!]
\centering
\scalebox{0.9}{
\begin{tabular}{ c|c c ccccccccccc } 
 \hline
$I$ & 1& 2& 3& 4& 5& $\hdots$& $\frac{n}{2^a}$& $\hdots$&$n-4$&$n-3$& $n-2$& $n-1$& $n$ \\
$a_i$ & 1& 4& 5& 8& 9& $\hdots$& $n$& $\hdots$&10&7& 6& 3& 2\\
 \hline
\end{tabular}}

\label{tab:Siegel-tukey weights}
\end{table}
\FloatBarrier
$^a$: If $n/2$ is odd, $i=\frac{n}{2}+1$ here. \\
If $n$ is  odd, the middle observation in the array is removed and the same weights are used for reduced $n$. In the case of ties these weights are adjusted by computing the mean of the weights for the tied observation. \textit{Refer to Table\ref{tab:table1} for illustration.} \\
The Siegel-Tukey test statistic is: $S_n=\sum_{i=1}^n a_iZ_i$,
where \begin{equation*}
  a_i =
    \begin{cases}
      2i & \text{for $i$ even, $1< i \leq n/2$}\\
      2i-1 & \text{for $i$ odd, $1\leq i \leq n/2$}\\
      2(n-i)+2 & \text{for $i$ even, $n/2< i \leq n$}\\
      2(n-i)+1 &  \text{for $i$ odd, $n/2< i \leq n$}  
    \end{cases}       
\end{equation*} 
The probability distribution of $S_n$ is same as that of the Wilcoxon rank-sum statistic, this will be used in computing the p-value. In our problem for testing for constant direction in the occurrence of an element, we compare the \textit{time-lag} vector with another vector of same length with all elements equal to median of the \textit{time-lag} vector. If \textit{lag} vector is more dispersed its ranks will be lower, as extreme values receive lower ranks, while the other group will receive more of the high scores assigned to the center. The p-value of the test if significant will indicate difference in scale and hence not a constant occurrence of the \textit{wins} while an insignificant p-value will indicate a constant direction of\textit{ wins}. 

\end{enumerate}
\subsection{Adjustment for Ties}
In our proposed approach, ties are a key factor, especially during the execution of the Siegel-Tukey test, as we compare the lag vector with another vector, in which all elements are equal to the median. 
The null distribution of the Siegel-Tukey test statistic may differ from that of the Wilcoxon-Mann-Whitney test when ties occur
A modification of the Siegel-Tukey test is proposed in \cite{vegelius} to resolve this possibility. When no ties occur, this modification reduces to the usual test. \\
\noindent \textbf{Vegelius's Method:}
Arrange the data in, say, ascending order. Begin from one side and let the most extreme value correspond to rank 1. If more than one element has this most extreme value, include all of them and give them the corresponding mid-rank. Let $n_1$ be the number of elements from side 1 ranked so far. continue with the ranking on the
other side and rank {\it enough elements} so that the number of elements included there is one more than on the first side. If ties occur at the latest element, include all elements having this value, even if this should increase the number of elements from this side. For the tied observations the mid-ranks (of consecutive numbers) should always be used. Let $n_2$ denote the number of elements ranked this far. Note that $n_2 \geq n_1+1$. Return to first side and rank so many elements that the total number ranked this far from this side is one more than $n_2$. If ties occur at the latest element we apply the same inclusion as above, which makes $n_1 \geq n_2+1$. 
Continue in this way from both sides until all elements are ranked. 
\\ \\
Consider the following illustration of this procedure:(Table\ref{tab:table1},Table\ref{tab2},Table\ref{tab3} )
\begin{center}
\begin{table}[hbt!]
\scalebox{0.9}{
\begin{tabular}{ c|c c cccccccccccccc } 
 \hline
Element No. & 1& 2& 3& 4& 5& 6& 7& 8& 9& 10& 11& 12& 13& 14& 15& 16 \\
Value &-3&-3&-1&-1&-1& 0& 0& 0& 1& 2& 2& 2& 3& 3& 3& 4\\
No tie rank& 1& 4& 5& 8& 9& 12& 13& 16& 15& 14& 11& 10& 7& 6 &3 &2\\
Tie rank & $2\frac{1}{2}$ &$2\frac{1}{2}$&$7\frac{1}{3}$&$7\frac{1}{3}$&$7\frac{1}{3}$&$13\frac{2}{3}$&$13\frac{2}{3}$&$13\frac{2}{3}$&15&$11\frac{2}{3}$&$11\frac{2}{3}$&$11\frac{2}{3}$&$5\frac{1}{3}$&$5\frac{1}{3}$&$5\frac{1}{3}$&2\\
 \hline
\end{tabular}}
\caption{Illustration of the Usual Procedure of Siegel-Tukey Test (Tied Case)}
\label{tab:table1}
\end{table}
\end{center}

\begin{center}
\begin{table}[hbt!]
\begin{tabular}{ c|c c cccccccccccccc } 
 \hline
Element No. & 1& 2& 3& 4& 5& 6& 7& 8& 9& 10& 11& 12& 13& 14& 15& 16 \\
Value &-3&-3&-1&-1&-1& 0& 0& 0& 1& 2& 2& 2& 3& 3& 3& 4\\
No tie rank& 1& 2& 7 &8 &9 &13& 14& 15 &16 &12& 11& 10& 6& 5 &4& 3\\
Tie rank & $1\frac{1}{2}$ &$1\frac{1}{2}$&8&8&8&14&14&14&16&11&11&11&5&5&5&3\\
 \hline
\end{tabular}
\caption{Illustration of the Vegelius Method for Siegel-Tukey Test (Tied Case)}
\label{tab2}
\end{table}
\end{center}

\begin{center}
\begin{table}[hbt!]
\scalebox{0.85}{
\begin{tabular}{ c|c c cccccccccccccc } 
 \hline
Element No. & 1& 2& 3& 4& 5& 6& 7& 8& 9& 10& 11& 12& 13& 14& 15& 16 \\
Value &-3&-2.5&-2&-1.5&0& 0.2& 0.7& 0.8& 1& 1.3& 1.5& 2& 3& 3.5& 4.3& 5\\
 \small{Rank (Usual Siegel-Tukey)} & 1& 4& 5& 8& 9& 12& 13& 16& 15& 14& 11& 10& 7& 6 &3 &2\\
Rank (Vegelius Method) &1& 4& 5& 8& 9& 12& 13& 16& 15& 14& 11& 10& 7& 6 &3 &2\\
 \hline
\end{tabular}}
\caption{Illustration of the Procedure in case of no ties}
\label{tab3}
\end{table}
\end{center}

The Vegelius correction has been adjusted during execution of the Siegel-Tukey test.  A function was written in R for this and  similar power comparisons as before executed.

\subsection{Proposed Approach}
This approach is designed to test the null hypothesis of randomness of occurrence of elements of a single type against the alternative of non-randomness or trend (increasing, decreasing or constant). It is based on the idea of ``pattern" and ``direction". The steps to be executed are as follows:

\begin{enumerate}[label=(\roman*)]
\item  The test will determine which type of element has higher frequency in the sample (1 or 0). Say there are more number of \textit{ones} than \textit{zeros} and number of ones is $k$.
\item Then it will find the positions of 1 in the ordered sequence, let the vector be $y_k$.
\item Define a lag vector: $y'$ of length $k-1$ where: 
$y'[i]= y[i+1]-y[i], 1\leq i \leq (k-1)$
\item Compute the mean of $y'$ and round up to the nearest integer value, say $\alpha$. 
\item Count the number of elements in $y'$ less than or equal to $\alpha$, say $x$.
\item Perform an exact binomial test with null proportion of $0.5$ and number of trials = k-1 and number of success = $x$.  A significant p-value will indicate any pattern in the occurrence of \textit{wins}. 
\item Define a vector $times$ of length $k-1$ with magnitudes increasing, let it be $times = (1,2,3,\hdots , (k-1))$.  Then  test for alternative hypothesis of trend(increasing, decreasing or constant),if any, is performed using Kendall Correlation Coefficient ($\tau$). A significant p-value will indicate the direction of \textit{wins}.
\item If the p-value of the above test is non significant, then define another vector $y''$ of length $k-1$ with all elements equal to the median of $y'$. We compare the lag vector $y'$ with $y''$ to check for any constant pattern. Now perform the ties-corrected method of  Siegel Tukey Test (procedure as mentioned in  (\cite{vegelius})). Note that the insignificant p-value will indicate a constant direction of \textit{wins}. 
\end{enumerate}

\subsection{Computing Size or Power of this Test}
The proposed test rejects the null hypothesis of randomness when there is evidence of either ``pattern" or ``direction" or both. That is, if the \textit{Binomial test component} shows a significant p-value then we conclude the presence of a possible pattern and thus non-randomness of the sequence, or if the Kendall Correlation test shows a significant p-value or the Siegel- Tukey test (with Vegelius correction) shows an insignificant p-value then we conclude the presence of possible increasing/decreasing or constant direction in the occurence of elements and reject the null hypothesis of randomness. 
\\ \\
\textbf{Testing the Illustrative Example}\\
Consider the binary sequence of 25 rounds of game described in Section \ref{sec:run_test_illustrative_example}. The usual Runs test for this sequence indicates the presence of randomness. 
We apply the proposed test to this sequence. 
\begin{enumerate}[label=(\roman*)]
\item The element $1$ has larger frequency with $n_1=16$. 
\item Finding the positions of the 1: $y_{16}= (1, 3 , 4 , 9 ,10 ,11 ,13 ,16 ,17, 18, 20, 21, 22 ,23, 24, 25)$
\item Lag vector: $y' =(3-1,4-3,9-4,\hdots, 24-23, 25-24) = (2 ,1 ,5 ,1 ,1 ,2 ,3 ,1 ,1 ,2 ,1, 1, 1, 1, 1)$
\item Mean of $y'$: $\frac{2+1+5+1+1+2+3+1+1+2+1+1+1+1+1}{15} = 1.6$ and rounding off to the nearest interest we set $\alpha =2$. 
\item Number of elements in $y'$ less than equal to $\alpha=2$: 2
\item Result when performing the exact binomial test: number of successes = 2, number of trials = 15, p-value = 0.007385
\\  As the p-value is significant for the Exact Binomial test, we can conclude that about the existence of pattern in the sequence and hence reject the null hypothesis of randomness.
\textit{The later two steps have been performed for ease of understanding of this test}.
\item The \textit{times} vector is: $( 1  ,2 , 3,  4,  5,  6,  7,  8,  9, 10, 11, 12, 13, 14, 15)$. The result for Kendall correlation test shows: tau = -0.349, 2-sided p-value =0.12001
\item As the p-value of the Kendall correlation test is significant, the Siegel-Tukey test with Vegelius correlation has been performed. \\
The median of $y'= 1$, and $y''=(1,1,1,\hdots,1,1)$ \{15 times\}. Comparing $y'$ and $y''$, the p-value of the Siegel- Tukey test (with Vegelius correction): 0.06307, which is insignificant indicating a constant direction of wins. This is also supported by Figure \ref{fig:runtest-example1}. 

\end{enumerate}
 \noindent {\bf Simulations and Results:}
Power comparisons were carried out to compare Runs Test and the proposed approach with sample size $n=20$ and $b = 0.01, 0.02, 0.03, 0.04, 0.05,$ $0.06, 0.07, 0.08, 0.09, 0.10$. In each case 1000 simulations were used to estimate the power.
\\To understand the performance of the proposed test, power estimation was performed for varying sample sizes ($n = 15, 20, 25, 30, 35$) and a finer scale level: $b$ = 0.0001, 0.0011, 0.0021, 0.0031, 0.0041, 0.0051, 0.0061, 0.0071, 0.0081, 0.0091, 0.0101. Again here in each case 1000 simulations were used to estimate the power.\\

\begin{figure}[h]
    \centering
    \begin{subfigure}[b]{0.5\textwidth}
         \centering
         \includegraphics[width=\textwidth]{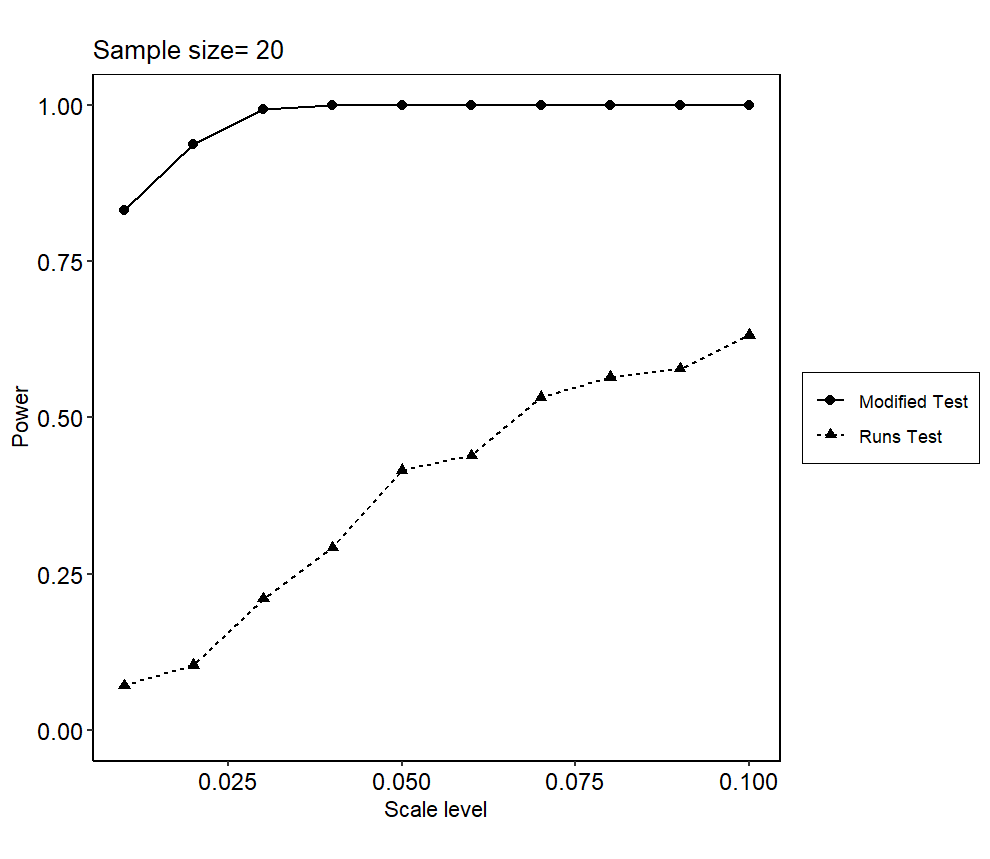}
         \caption{Comparison with Runs Test}
         \label{fig:modified vs runs}
     \end{subfigure}
     \hfill
     \begin{subfigure}[b]{0.45\textwidth}
         \centering
         \includegraphics[width=\textwidth]{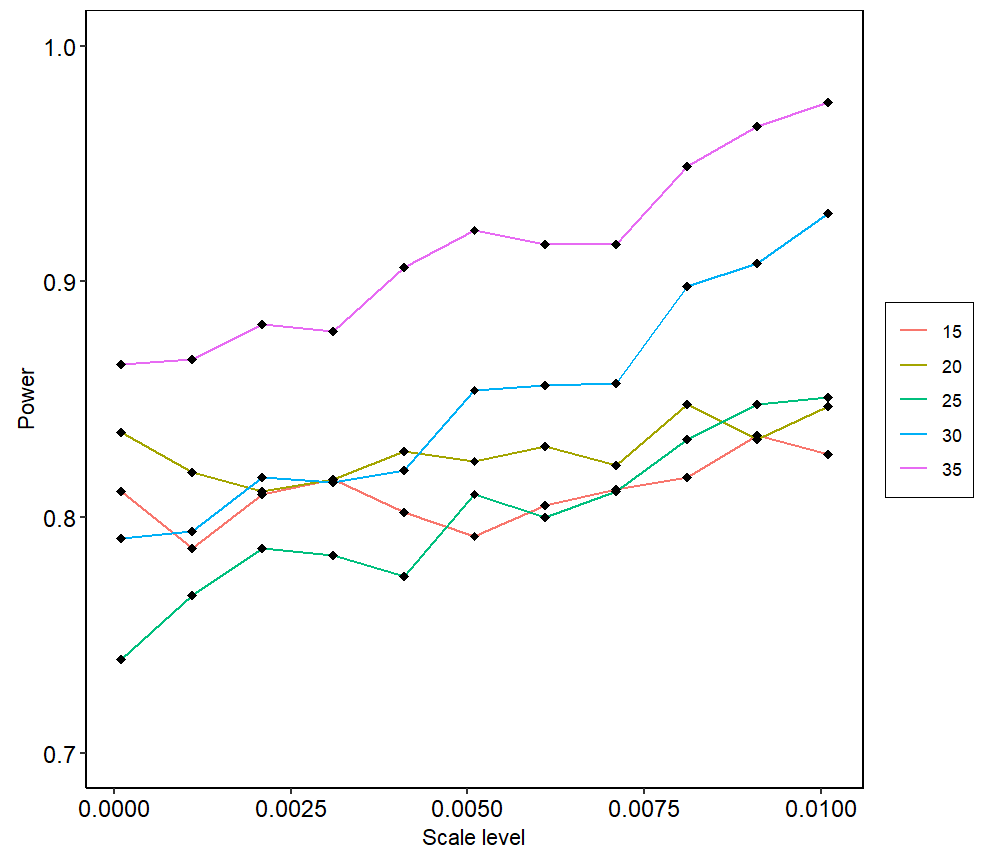}
         \caption{Comparisons over sample sizes}
         \label{fig:modified sample sizes compare}
     \end{subfigure}
    \caption{Power comparisons from Proposed Test(with Vegelius Correction)}
    \label{fig:modified test results}
\end{figure}
\FloatBarrier
Observe (Figure \ref{fig:modified vs runs}) that the power of the proposed test is much higher than the Runs Test and the power quickly increases to one. \\

The power of the proposed approach has been compared with that of the conventional runs test. A major point to be noted is that under the Runs test, the null hypothesis is randomness of the entire sequence \textit{(consisting of 1s and 0s)} however in the proposed test, the null hypothesis is randomness of occurrence of element of a single type \textit{(say, 1s or 0s)}. The reason for comparison between the two tests is to motivate the reader into the significance of the proposed method. 
Figure \ref{fig:modified sample sizes compare} clearly indicates that estimated power tends to increase in general for increasing $b$ and for larger sample sizes. 
\textit{Appendix B shows similar power comparisons when the Vegelius correction has not been incorporated.}
\section{Conclusion}
Given an ordered binary sequence of wins and losses over time, and with the objective of determining randomness in the occurrence of wins (or losses), the Runs Test may yield results that contradict the intuition suggested by scatter plots of win proportions over time. This paper demonstrates that it is possible to design a test suitable for this purpose by computing the gaps between consecutive wins(or losses) and applying exact binomial tests, along with non-parametric tests such as Kendall's Tau and the Siegel-Tukey test (1960) for scale problems, to determine heteroscedastic patterns and the direction of win occurrences. Further modifications suggested by Vegelius (1982) have also been applied to the Siegel-Tukey test (1960) to achieve more accurate results. We obtain very good power against plausible alternatives in simulation for both the formulations. 

Although the results are demonstrated in the context of win-loss sequences, the proposed methodology can be applied in various other contexts where testing for similar patterns is important in any dichotomous data. We hope that this test will prove useful for practitioners analyzing categorical data.
\section*{Acknowledgement}
I am  grateful to Prof. Diganta Mukherjee, Indian Statistical Institute for his valuable guidance and suggestions that greatly improved the article.

\newpage
\section*{Appendix A}
\label{appendix}
\begin{enumerate}
    \item Claim: Under the condition that in a binary sequence, frequency of type 1 element is greater than the frequency of type 2 element, if there exists a pattern in the occurrence of type 1 element then the distances between any two consecutive type 1 elements, that is, elements of $y'$, will be relatively ``similar" in general.
    \\
    \textit{Proof of the Claim}: 
Let the frequency of type 1 elements be $n_1$ and that of type 2 elements be $n_2$. Assume $n_1>1$ and $n_2>1$. 
Thus $n_1+n_2=n$ (total number of elements in the sequence). Let $K$ denote the sum of elements of $y'$. Note that by definition the number of elements in $y'$ is $(n_1-1)$.Then for some fixed constant 
$c \geq 0$: $n_2 = K-(n_1-1)+c $.  Clearly $c$ is a whole number.
If $K = (k-1)(n_1-1)$ for some constant $k$. Then $n_1-n_2 = n_1 -(k-1)(n_1-1)+n_1-1-c = (3-k)(n_1-1)+1-c$. \\
For $n_1>n_2$, we must have $(3-k)(n_1-1) > c-1$.\\
Consider the sub-case: 

\begin{enumerate}[label=(\roman*)]
\item $c \geq 1$: Then we must have $k \leq 3$.
\item $c=0$: Then $(3-k)(n_1-1) > -1$, the left side of the inequality is product of two integers and $n_1-1\geq 1$, thus $k\leq3$.
\end{enumerate}

Thus the mean of elements of $y'$ : $(k-1)$ will be less than or equal to 2, and the elements of $y'$ will be natural numbers thus indicating that the distances between any 2 consecutive type 1 elements will be similar, thereby proving the claim.
\end{enumerate}
\newpage
\section*{Appendix B}
\begin{figure}[h]
    \centering
    \begin{subfigure}[b]{0.5\textwidth}
         \centering
         \includegraphics[width=\textwidth]{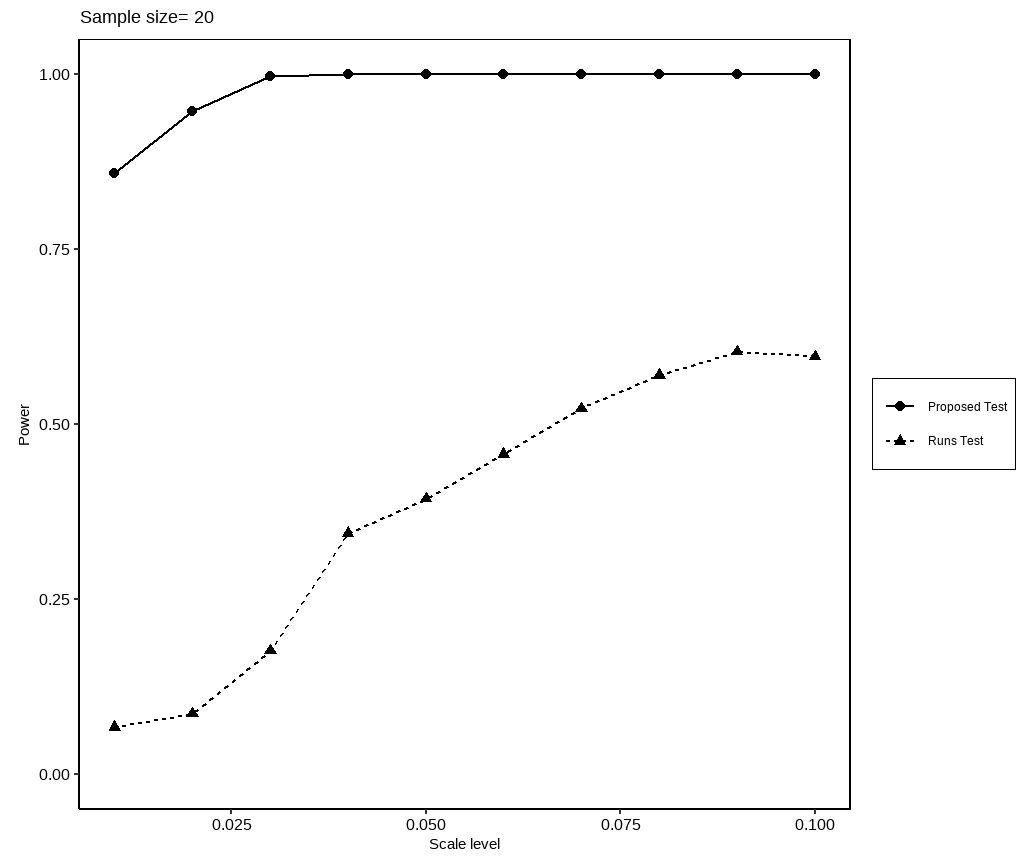}
         \caption{Comparison with Runs Test}
         \label{fig:comparison1}
     \end{subfigure}
     \hfill
     \begin{subfigure}[b]{0.45\textwidth}
         \centering
         \includegraphics[width=\textwidth]{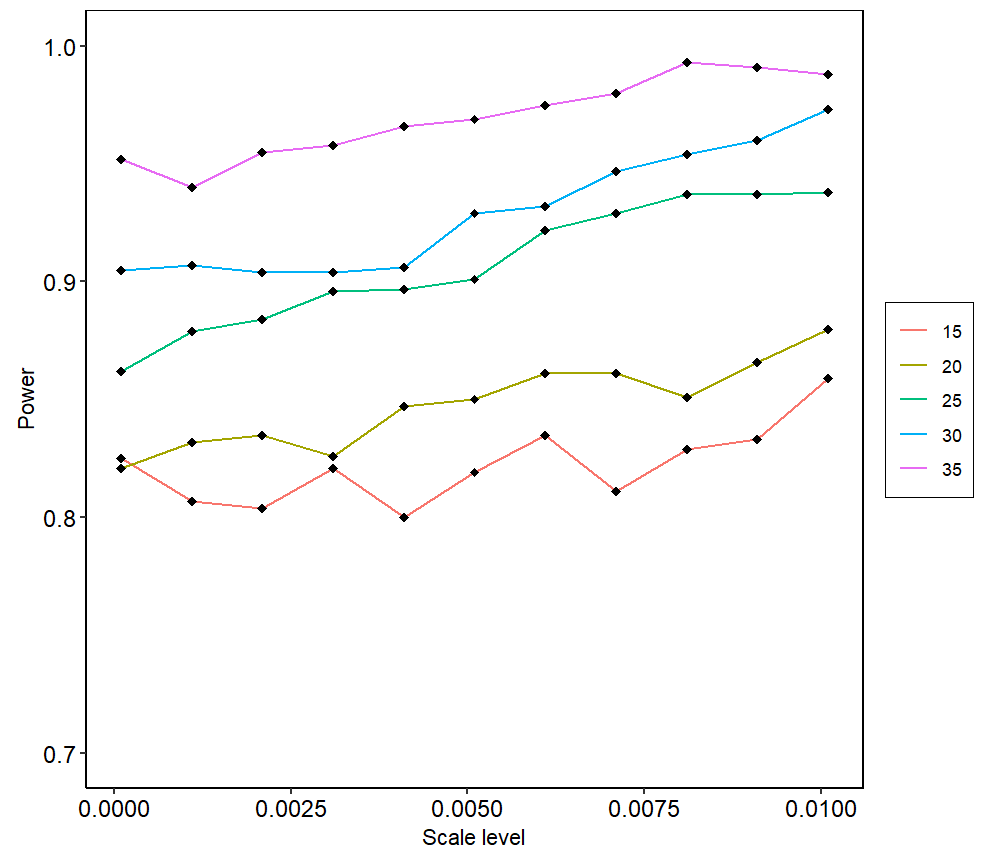}
         \caption{Comparisons over sample sizes}
         \label{fig:comparison2}
     \end{subfigure}
    \caption{Power comparisons from Proposed Test (without Vegelius Correction)}
    \label{fig:proposed test results}
\end{figure}

\textbf{Power Comparisons Without Vegelius Correction}\\
\\ We consider the power simulation results when the corrections by \cite{vegelius} are not incorporated. 
The power performance is shown in Figure \ref{fig:comparison1}, which is similar to Figure \ref{fig:modified vs runs} , and estimated power of the proposed test (without Vegelius correction) is also much higher than that of the Runs Test. 
Additionally, to understand the performance of this test, power estimation was performed for varying sample sizes ($n = 15, 20, 25, 30, 35$) and a finer scale level: $b$ = 0.0001, 0.0011, 0.0021, 0.0031, 0.0041, 0.0051, 0.0061, 0.0071, 0.0081, 0.0091, 0.0101. 1000 simulations were used. 
\FloatBarrier

Figure \ref{fig:proposed test results} clearly indicates that the estimated power tends to increase in general for increasing $b$ and for larger sample sizes. 

Comparing Figures \ref{fig:modified sample sizes compare} and \ref{fig:comparison2}, we observe that the power obtained in the proposed test correcting for Vegelius's method is slightly lower than the original proposal, which may be considered the cost of adjustment (for ties). However, the power reaches 1 faster (as the alternative moves further away from the null) in the Vegelius-corrected proposal than in the uncorrected proposal. Thus, both versions perform comparably well in their separate domains of usefulness (the cases of no ties and with ties respectively).

\end{document}